%
\input epsf
%
%
%
\def\today{\ifcase\month\or January\or February\or March\or April\or May\or
June\or July\or August\or September\or October\or November\or December\fi
\space\number\day, \number\year}
%
%
\newcount\notenumber

\def\note{\global\advance\notenumber by 1 \footnote{$^{\the\notenumber}$}}
%
%
\def\alphasecnumber{{\rm\ifcase\secnumber\or A\or B\or C\or D\or E\or F\or
G\fi}}
\newif\ifsectionnumbering
\newcount\eqnumber
\def\cleareqnumber{\eqnumber=0}
\def\numbereq{\global\advance\eqnumber by 1
\ifinappendix
  \ifsectionnumbering\eqno(\alphasecnumber.\the\eqnumber)\else
  \eqno(\the\eqnumber)\fi
\else
  \ifsectionnumbering\eqno(\the\secnumber.\the\eqnumber)\else
  \eqno(\the\eqnumber)\fi
\fi}
\def\eqalinno{{\global\advance\eqnumber by 1}
\ifsectionnumbering(\the\secnumber.\the\eqnumber)\else(\the\eqnumber)\fi}
\def\name#1{
  \ifinappendix
    \ifsectionnumbering\xdef#1{\alphasecnumber.\the\eqnumber}
      \else\xdef#1{\the\eqnumber}\fi
  \else
    \ifsectionnumbering\xdef#1{\the\secnumber.\the\eqnumber}
      \else\xdef#1{\the\eqnumber}\fi
\fi}

\sectionnumberingtrue
%
%
\newcount\refnumber

\immediate\openout1=refs.tex
\immediate\write1{\noexpand\frenchspacing}
\immediate\write1{\parskip=0pt}
\def\ref#1#2{\global\advance\refnumber by 1%
[\the\refnumber]\xdef#1{\the\refnumber}%
\immediate\write1{\noexpand\item{[#1]}#2}}
\def\tie{\noexpand~}

%
%
\font\twelvebf=cmbx10 scaled \magstep1
\newcount\secnumber

\def\newsection#1.{\ifsectionnumbering\cleareqnumber\else\fi%
	\global\advance\secnumber by 1%
	\bigbreak\bigskip\par%
	\line{\twelvebf \the\secnumber. #1.\hfil}\nobreak\medskip\par\noindent}
%
%
%
\def \sqr#1#2{{\vcenter{\vbox{\hrule height.#2pt
	\hbox{\vrule width.#2pt height#1pt \kern#1pt
		\vrule width.#2pt}
		\hrule height.#2pt}}}}

%
%
%
\newdimen\fullhsize
\def\fiddle{\fullhsize=6.5truein \hsize=3.2truein}
\def\fullline{\hbox to\fullhsize}
\def\mkhdline{\vbox to 0pt{\vskip-22.5pt
	\fullline{\vbox to8.5pt{}\the\headline}\vss}\nointerlineskip}
\def\mkftline{\baselineskip=24pt\fullline{\the\footline}}
\let\lr=L \newbox\leftcolumn
\def\twocolumns{\fiddle
	\output={\if L\lr \global\setbox\leftcolumn=\columnbox
		\global\let\lr=R \else \doubleformat \global\let\lr=L\fi
		\ifnum\outputpenalty>-20000 \else\dosupereject\fi}}
\def\doubleformat{\shipout\vbox{\mkhdline
		\fullline{\box\leftcolumn\hfil\columnbox}
		\mkftline} \advancepageno}
\def\columnbox{\leftline{\pagebody}}
\magnification=1200
\def\pr#1 {Phys. Rev. {\bf D#1},\tie}
\def\pe#1 {Phys. Rev. {\bf #1\tie}}
\def\pre#1 {Phys. Rep. {\bf #1\tie}}
\def\pl#1 {Phys. Lett. {\bf B#1},\tie}
\def\prl#1 {Phys. Rev. Lett. {\bf #1},\tie}
\def\np#1 {Nucl. Phys. {\bf B#1},\tie}
\def\ap#1 {Ann. Phys. (NY) {\bf #1\tie }}
\def\cmp#1 {Commun. Math. Phys. {\bf #1\tie }}
\def\imp#1 {Int. Jour. Mod. Phys. {\bf A#1\tie }}
\def\mpl#1 {Mod. Phys. Lett. {\bf A#1\tie}}
\def\jhep#1 {JHEP {\bf #1},\tie}
\def\nuo#1 {Nuovo Cimento {\bf B#1\tie}}
\def\class#1 {Class. Quant. Grav. {\bf #1},\tie}
\def\for#1 {Fortsch. Phys. {\bf #1},\tie}
\def\tie{\noexpand~}
\def\ov{\bar}

\parskip=15pt plus 4pt minus 3pt
\headline{\ifnum \pageno>1\it\hfil The Radion in
the Karch-Randall Braneworld\else \hfil\fi}
\font\title=cmbx10 scaled\magstep1
\font\tit=cmti10 scaled\magstep1
\footline{\ifnum \pageno>1 \hfil \folio \hfil \else
\hfil\fi}
\raggedbottom


\overfullrule0pt

\def\ft#1#2{{\textstyle{{#1}\over{#2}}}}
\def\fft#1#2{{{#1}\over{#2}}}
\def\sgn{{\,\rm sgn}}

\newif\ifinappendix\inappendixfalse
\def\newappendix#1.{\ifinappendix\else\secnumber=0\inappendixtrue\fi%
        \ifsectionnumbering\cleareqnumber\else\fi%
	\global\advance\secnumber by 1%
	\bigbreak\bigskip\par%
	\line{\twelvebf #1.\hfil}\nobreak\medskip\par\noindent}
\def\lref#1#2{\global\advance\refnumber by 1%
\xdef#1{\the\refnumber}%
\immediate\write1{\noexpand\item{[#1]}#2}}
\def\rtitle#1{{\sl #1},}
\font\btitle=cmbx12 scaled\magstep1


\rightline{\vbox{\hbox{RU03-05-B}\hbox{MCTP-03-47}\hbox{hep-th/0311216}}}
\vfill
\centerline{\btitle The Radion in the Karch-Randall Braneworld}
\vskip4em
{\centerline{\title Ioannis Giannakis${}^{a}$, James T. Liu${}^{b}$
and Hai-cang Ren${}^{a}$ \footnote{$^{\dag}$}
{\rm e-mail: \vtop{\baselineskip12pt
\hbox{giannak@summit.rockefeller.edu, jimliu@umich.edu,}
\hbox{ren@summit.rockefeller.edu}}}}}
\bigskip
\centerline{$^{(a)}${\tit Physics Department, The Rockefeller University}}
\centerline{\tit 1230 York Avenue, New York, NY
10021--6399}
\bigskip
\centerline{$^{(b)}${\tit Michigan Center for Theoretical Physics}}
\centerline{\tit Randall Laboratory, Department of Physics,
University of Michigan}
\centerline{\tit Ann Arbor, MI 48109--1120} 
\vskip4em
\centerline{\title Abstract}
\bigskip
{\narrower\narrower\noindent
In a braneworld context, the radion is a massless mode coupling to the
trace of the matter stress tensor.  Since the radion also governs the
separation between branes, it is expected to decouple from the physical
spectrum in single brane scenarios, such as the one-brane Randall-Sundrum
model.  However, contrary to expectations, we demonstrate that the
Karch-Randall radion always remains as a physical excitation, even in
the single brane case.  Here, the radion measures the distance not
between branes, but rather between the brane and the anti-de Sitter
boundary on the other side of the bulk.}
\vfill\break

\newsection Introduction.%
The idea that gravity may be trapped on a braneworld with an infinite extra
dimension has led to much recent excitement among both particle physicists
and cosmologists.  From a Kaluza-Klein perspective, it has long been thought
that a non-compact extra dimension would lead to a continuous spectrum
without any localized gravity.  However, as demonstrated by Randall and
Sundrum
\lref{\randall}{L. Randall and R. Sundrum,
\rtitle{A large mass hierarchy from a small extra dimension}
Phys. Rev. Lett. {\bf83}, 3370 (1999) [hep-ph/9905221].}%
\lref{\sundrum}{L. Randall and R. Sundrum,
\rtitle{An alternative to compactification}
Phys. Rev. Lett. {\bf83}, 4690 (1999) [hep-th/9906064].}%
[\randall,\sundrum],
the traditional Kaluza-Klein result may be evaded by using a warped
compactification of the form
$$
ds^2=e^{2A(y)}\bar g_{\mu\nu}(x)dx^\mu dx^\nu+dy^2.
\numbereq\name{\eqAdS5}
$$
For the Randall-Sundrum braneworld, the warp factor is given by
$$
e^{2A}=e^{-2\kappa|y|},
\numbereq\name{\eqRS}
$$
while the brane metric is flat, $\bar g_{\mu\nu}=\eta_{\mu\nu}$.  This
choice ensures that the bulk metric is simply that of AdS$_5$, written
in horospherical coordinates, and with cosmological constant
${\cal R}_{MN}=-4\kappa^2G_{MN}$.  The original Randall-Sundrum model
[\randall], denoted RS1, consisted of a compact extra dimension
sandwiched between two flat branes, located at $y=0$ (which we denote
the physical brane) and at $y=y_1$ (which we denote the regulator brane).
In other words, the range of $y$ in (\eqRS)
is restricted to $y\in[0,y_1]$.  However, in a subsequent modification,
the second brane was removed ($y_1\to\infty$), yielding a one brane model
[\sundrum], denoted RS2.  In both models, the graviton is trapped at
the kink in the warp factor at $y=0$ ({\it i.e.}~on the physical brane).
However it is in the latter that the novel feature of graviton localization
with an infinite extra dimension shows up.

Subsequently, it was realized that the fine tuning of brane tensions inherent
in the Randall-Sundrum model may be relaxed without destroying the graviton
trapping feature.  Gravity in the resulting bent braneworld model was studied
by Karch and Randall
\ref{\karch}{A. Karch and L. Randall, \jhep0105 (2001) 008, hep-th/0105108.},
which showed that a massive graviton may be trapped on an AdS$_4$ braneworld,
while a massless one would be trapped on a dS$_4$ braneworld.  The former
massive graviton has been the object of much investigation, particularly
in regards to avoidance of the van Dam-Veltman-Zakharov discontinuity
\lref{\vdv}{H. van Dam and M. Veltman,
\rtitle{Massive and massless Yang-Mills and gravitational fields}
Nucl.\ Phys.\ B {\bf 22}, 397 (1970).}%
\lref{\zak}{V.I. Zakharov, JETP Lett. {\bf 12}, 312 (1970).}%
[\vdv,\zak].
In particular, it was demonstrated in
\lref{\Kogan}{I.I. Kogan, S. Mouslopoulos and A. Papazoglou,
\rtitle{The $m \to 0$ limit for massive graviton in $dS_4$ and $AdS_4$:
How to circumvent the van Dam-Veltman-Zakharov discontinuity}
Phys.\ Lett.\ B {\bf 503}, 173 (2001) [hep-th/0011138].}%
\lref{\Porrati}{M. Porrati,
\rtitle{No van Dam-Veltman-Zakharov discontinuity in AdS space}
Phys.\ Lett.\ B {\bf 498}, 92 (2001) [hep-th/0011152].}
[\Kogan,\Porrati]
that the spin-2 discontinuity is absent in an AdS background.  Similarly,
a possible spin-3/2 discontinuity is also absent
\lref{\vanN}{P.A. Grassi and P. van Nieuwenhuizen,
\rtitle{No van Dam-Veltman-Zakharov discontinuity for supergravity in
AdS space} Phys.\ Lett.\ B {\bf 499}, 174 (2001) [hep-th/0011278].}%
\lref{\Deser}{S. Deser and A. Waldron,
\rtitle{Discontinuities of massless limits in spin 3/2
mediated interactions and cosmological supergravity}
Phys.\ Lett.\ B {\bf 501}, 134 (2001) [hep-th/0012014].}%
[\vanN,\Deser],
thus avoiding a possible obstruction to the supersymmetrization of the model.
Some novel features remain, however, and indicated by the presence of a
quantum discontinuity
\lref{\Dilkes}{F.A. Dilkes, M.J. Duff, J.T. Liu and H. Sati,
\rtitle{Quantum discontinuity between zero and infinitesimal graviton
mass with a $\Lambda$ term} Phys.\ Rev.\ Lett.\  {\bf 87}, 041301 (2001)
[hep-th/0102093].}%
\lref{\Duff}{M.J. Duff, J.T. Liu and H. Sati,
\rtitle{Quantum discontinuity for massive spin 3/2 with a $\Lambda$ term}
hep-th/0211183.}%
[\Dilkes,\Duff],
as well as a natural AdS-Higgs mechanism for generation of graviton mass
\lref{\Porrati}{M. Porrati,
\rtitle{Higgs phenomenon for 4-$D$ gravity in anti de Sitter space}
JHEP {\bf 0204}, 058 (2002) [hep-th/0112166].}%
\lref{\Duffkr}{M.J. Duff, J.T. Liu and H. Sati,
\rtitle{Complementarity of the Maldacena and Karch-Randall pictures}
AIP Conf.\ Proc.\  {\bf 655}, 155 (2003)
[hep-th/0207003].}%
\lref{\Porratia}{M. Porrati,
\rtitle{Higgs phenomenon for the graviton in AdS space}
Mod.\ Phys.\ Lett.\ A {\bf 18}, 1793 (2003) [hep-th/0306253].}%
[\Porrati,\Duffkr,\Porratia].

For the AdS$_4/$AdS$_5$ braneworld, with four-dimensional cosmological
constant $\bar R_{\mu\nu}=-3k^2\bar g_{\mu\nu}$, the warp factor $e^{2A}$
of (\eqAdS5) takes the form 
$$
e^{2A}={k^2\over \kappa^2}\cosh^2\kappa(|y|-\bar y),
\numbereq\name{\eqKR}
$$
where the constant $\bar y$ is determined by continuity across the brane
and the normalization requirement that the warp factor on the physical
brane at $y=0$ is one, {\it i.e}
$$
{\rm sech}^2\kappa\bar y={k^2\over\kappa^2}.
\numbereq\name{\eqconvent}
$$
In analogy with the Randall-Sundrum model, we
shall denote by KR1 the Karch-Randall model [\karch] which consists
of two AdS$_4$ branes embedded in AdS$_5$ (with the physical brane at
$y=0$ and the regulator brane at $y=y_1$).  Similarly, we will use KR2
to denote the model in which the second brane is sent to infinity.

While much attention has been paid to the trapping of a massless graviton
on the Randall-Sundrum brane (and a massive one on the Karch-Randall brane),
perhaps of no less importance for the braneworld is the possible existence
of a radion (spin-0) degree of freedom.  Of course, the radion has a natural
origin from a Kaluza-Klein point of view.  After all, the five-dimensional
graviton naturally decomposes into a set of spin-0, spin-1 and spin-2 modes
in four dimensions.  By imposing $Z_2$ boundary conditions, the massless
spin-1 degree of freedom may be projected out.  Alternatively, the warped
geometry no longer has an isometry generated by $\partial/\partial y$;
hence the corresponding graviphoton ought to be absent from the spectrum.
However, the radion may in principle survive.

A linearized analysis of both RS1
\ref{\charmousis}{C. Charmousis, R. Gregory and V. Rubakov,
\rtitle{Wave Function of the Radion in a Brane World}
Phys. Rev. {\bf D62}, 067505 (2000) [hep-th/9912160].},
and KR1
\ref{\chacko}{Z. Chacko and P. Fox,
\rtitle{Wave Function of the Radion in the DS and ADS  Brane World}
Phys. Rev. {\bf D64}, 024015 (2001) [hep-th/0102023].},
indicates that the radion is
in fact present in their spectra.  The natural physical picture is that
the radion parametrizes the distance between the two branes (physical and
regulator).  Hence the radion mode corresponds to fluctuations of the
relative position between the branes, and is a scalar which couples to the
trace of the stress energy tensors on the branes.  Along with this picture,
it is clear that as we push the second brane in the RS1 model to infinity,
the radion mode blows up and in the limit becomes non-normalizable.  Hence
it is no longer present in the RS2 model, which has only the physical brane
at $y=0$.  Another way to understand this is that without a second brane,
there is no longer a physically relevant distance that may be measured by
the radion.  Any shift in $y$ in the RS2 model may be compensated for by a
Weyl rescaling of the four-dimensional brane metric.

On the other hand, the physics of removing the second brane in the KR1 model
is different from that of the Randall-Sundrum model.  In the Randall-Sundrum
case, the limit $y\to\infty$ corresponds to a Cauchy horizon of AdS$_5$,
as is evident from (\eqRS).  Thus pushing the second brane to infinity in
RS1 corresponds to pushing it off to a horizon.  On the other hand, as may
be seen from (\eqKR), $y\to\infty$ in the Karch-Randall model corresponds to
a blowing up warp factor, and hence to the (partial) boundary of AdS$_5$.
In pushing the second KR1 brane to infinity, one is no longer making it
disappear behind a horizon, but instead is placing it at the AdS$_5$
boundary, where it remains visible to massless particles.  As a result, the
radion remains in the physical spectrum in KR2, and now measures the
distance between the physical brane and the AdS$_5$ boundary.

To demonstrate the physical difference between the RS1 and KR1 limits, we
may compute the time it takes for light to travel from the physical brane
($y=0$) to the regulator brane ($y=y_1$), and back.
For the KR1 model, we find that
$$
t=2{\int_{0}^{y_1}}dy\,e^{-A}=
{4\over k}[\tan^{-1} e^{\kappa(y_1-\bar y)}-
\tan^{-1} e^{-\kappa\bar y}]
\numbereq\name{\eqxano}
$$
As we take the second brane to infinity, $y_1 \to \infty$,
the time $t$ remains finite, $t=\fft4k\tan^{-1}e^{\kappa\bar y}$.
Thus we observe that in the KR1 model the second brane can never be removed,
even when it is pushed to infinity.  Note that the Randall-Sundrum
braneworld can be recovered from the Karch-Randall one by taking the
limit $k \to 0$ while keeping ${k\over {\kappa}}{\cosh{{\kappa}\bar y}}=1$
fixed.  In this case, we find that the light travel time in RS1 is
$t={2\over {\kappa}}(e^{{\kappa}y_1}-1)$. In contrast to the KR1 model,
this expression diverges as we take the limit $y_1 \to \infty$, confirming
the physical distinction between the boundary and Cauchy horizon of AdS$_5$
in the braneworld context.

What we have argued above on physical grounds is that the Karch-Randall
radion is present in both the KR1 and KR2 models.  In this paper we shall
demonstrate that this is true by careful examination of the limit that
the $y$-coordinate of the regulator brane is taken to infinity, namely
$y_1\to\infty$.  We will examine both the KR1 and RS1 models in the absence
of matter on the regulator brane.  In the case of KR1, we will demonstrate
that the radion mode is retained in the limit, independent of whether
Newmann or Dirichlet boundary conditions are imposed on the regulator brane.
Furthermore, we will show that the linearized gravity analysis goes smoothly
to that of KR2 with a natural boundary condition at $y=\infty$.  On the
other hand, we will see that the radion mode in RS1 decouples in the
$y_1\to\infty$ limit, and that the linearized gravity goes smoothly to that
of RS1.

This paper is organized as follows.  In section 2, we derive
the linearized Einstein equations of the metric fluctuations
about the braneworld, investigate the gauge symmetries (diffeomorphisms)
that respect the standard (Gauss-normal) form of the metric,
and discuss the possible boundary conditions on the branes.
In section 3, we investigate the transverse-traceless modes,
paying particular attention to boundary conditions for the two brane 
scenario.  We also demonstrate explicitly how to recover the one brane
scenario by pushing the regulator brane to infinity.  We then
explore the details of the radion mode in section 4,
especially the KR2 limit of the KR1 brane model. The RS1 limit of KR1,
the RS2 limit of RS1 and the RS2 limit of KR2 will be
discussed in section 5.

\newsection Linearized gravity analysis.%
A common feature to both the Randall-Sundrum and Karch-Randall models is
the presence of an AdS$_5$ bulk geometry.  The main difference between the
models then arises in how the 3-branes are embedded in the bulk.  In
particular, the flat braneworld model arises only with a fine tuning of
the brane tensions, while the AdS$_4$ braneworld corresponds to a reduced
brane tension, in effect providing insufficient brane vacuum energy to
compensate for the negative bulk cosmological constant.

Since we are interested in the metric fluctuations in the braneworld, we
take as our starting point the general five-dimensional metric
$$
ds^2=G_{MN}(X)dX^MdX^N.
\numbereq\name{\eq5dmetric}
$$
The action is simply that of gravity with a bulk cosmological constant
coupled to the physical and regulator branes
$$
S=K\int d^5X\sqrt{-G}({\cal R}+12\kappa^2)+S^{(0)}+S^{(1)},
\numbereq\name{\eqEH}
$$
where ${\cal R}$ is the scalar curvature computed from the five dimensional
metric $G_{MN}$ and $K=1/16\pi G_5$.  The terms $S^{(0)}$ and $S^{(1)}$
represent the matter actions on the physical and regulator branes,
respectively.  The Einstein equation arising from (\eqEH) is 
$$
{\cal G}_{MN}-6\kappa^2G_{MN}=8\pi G_5(T_{MN}^{(0)}+T_{MN}^{(1)}),
\numbereq\name{\eqeins}
$$
where ${\cal G}_{MN}={\cal R}_{MN}-\fft12G_{MN}{\cal R}$ is the
five-dimensional Einstein tensor and $T_{MN}^{(s)}$ is the stress tensor 
of either the physical brane ($s=0$) or the regulator brane ($s=1$).

In order to highlight the physics of the branes and their embedding in
the bulk, it is useful to make a gauge choice for the metric $G_{MN}$
that splits it into components along the brane and perpendicular to it.
In particular, we often consider a Gauss-normal coordinate system
$$
ds^2=g_{\mu\nu}(x,y)dx^\mu dx^\nu+dy^2,
\numbereq\name{\eqGN}
$$
where $y$ is the direction perpendicular to the brane.  For a one brane
scenario, the brane (say at $y=0$) remains straight with respect to this
gauge choice.  Note, however, that in a two brane model, it is not always
possible to keep both branes straight in a single coordinate patch.
Furthermore, the r\^ole that brane bending plays will also be important for
the matter coupling of the braneworld.  Of course, for the two brane case,
it is always possible to cover the physical space between the branes
using multiple coordinate patches.  For a coordinate system which is
straight with respect to the physical brane at $y=0$, the brane stress
tensor has the form
$$
T_{\mu\nu}^{(0)}=(\lambda_0g_{\mu\nu}+T_{\mu\nu})\delta(y),
\numbereq\name{\eqphy}
$$
where $\lambda_0={3\over {4{\pi}G_5}}{\tanh{{\kappa}{\ov y}}}$
is the brane tension
and $T_{\mu\nu}$ is
the stress tensor for additional matter on the brane.  On the other hand,
the regulator brane stress tensor is given by
$$
T_{\mu\nu}^{(1)}=\lambda_1g_{\mu\nu}\delta(y-y_1),
\numbereq\name{\eqreg}
$$
where $\lambda_1={3\over {4{\pi}G_5}}
e^{2A(y_1)}{\tanh{\kappa(y_1-{\ov y})}}$ and
the coordinate system is taken to be  straight with respect to the
regulator brane at $y=y_1$.  Note that we do not include matter on the
regulator brane.  In both cases, the stress energy is restricted to lie
on the brane directions, so that $T_{\mu 4}^{(s)}=T_{44}^{(s)}=0$.  Two
coordinate patches (one for the vicinity of each brane), and a single set
of transition functions ought to be sufficient for the two brane scenario.

We now turn to a linearized gravity analysis of the braneworld.  Writing
$G_{MN}=G_{MN}^{(0)}+\delta G_{MN}$, and expanding about a solution to
the Einstein equations, (\eqeins), we find that the action quadratic
in metric fluctuations is given by
$$
\Delta S=-{K\over 2}\int d^5X\sqrt{-G^{(0)}}
\Big[\delta G^{MN}(\delta{\cal G}_{MN}-6\kappa^2
\delta G_{MN})-8\pi G_5\delta g^{\mu\nu}T_{\mu\nu}\delta(y)\Big].
\numbereq\name{\eqfluc}
$$
To address the braneworld scenario, we consider metric fluctuations of
the Gauss-normal form
$$
ds^2=e^{2A(y)}(\bar g_{\mu\nu}(x)+h_{\mu\nu}(x,y))dx^\mu dx^\nu+dy^2.
\numbereq\name{\eqds}
$$
A computation of the Ricci tensor and the Ricci scalar to first order in
the fluctuation $h_{\mu\nu}$ yields
$$
\eqalign{
{\cal R}_{\mu\nu}&=-4\kappa^2e^{2A}(\bar g_{\mu\nu}+h_{\mu\nu})
+\ft12\bar g^{\rho\lambda}(\nabla_\mu\nabla_\rho h_{\lambda\nu}
+\nabla_\nu\nabla_\rho h_{\lambda\mu})
-\ft12(\nabla^2h_{\mu\nu}+\nabla_\mu\nabla_\nu h)\cr
&\kern2truecm-\ft12e^{2A}(\ddot h_{\mu\nu}+4\dot A\dot h_{\mu\nu}
+\dot A\bar g_{\mu\nu}\dot h)-k^2(h_{\mu\nu}-\bar g_{\mu\nu} h),\cr
{\cal R}_{4\rho}&=\ft12\bar g^{\nu\lambda}
(\nabla_\lambda\dot h_{\nu\rho}-\nabla_\rho\dot h_{\nu\lambda}),\cr
{\cal R}_{44}&=-4\kappa^2-\ft12\ddot h-\dot A{\dot h}, \cr
{\cal R}_{\hphantom{44}}&=-12k^2e^{-2A}-8{\ddot A}-20{\dot A}^2
-{\ddot h}-5{\dot A}{\dot h}+3k^2h
-{\nabla^2}h+{\nabla^\mu}{\nabla^\nu}h_{\mu\nu}.\cr}
\numbereq\name{\eqricci}
$$
Note that the four-dimensional indices $\mu,\nu,\ldots$ are raised and
lowered using the brane metric $\bar g_{\mu\nu}$.  Likewise, the covariant
derivatives are with respect to this metric, and $h=\bar g^{\mu\nu}
h_{\mu\nu}$.  Here the dots denote derivatives with respect to the transverse
coordinate $y$.

The Einstein equations, written in Ricci form, are given by
${\cal R}_{MN}+4{\kappa}^2g_{MN}=8{\pi}G_5
(T_{MN}-{1\over 3}Tg_{MN}){\delta}(y)$, and become
$$
\eqalign{
&{\nabla^2}h_{\mu\nu}-{\bar g}^{\lambda\rho}
({\nabla_\nu}{\nabla_\lambda}h_{\mu\rho}+{\nabla_\mu}
{\nabla_\lambda}h_{\rho\nu})+{\nabla_\mu}{\nabla_\nu}h
+2k^2(h_{\mu\nu}-\bar g_{\mu\nu}h)\cr
&\kern1.5truein+e^{2A}(\ddot h_{\mu\nu}+4\dot A\dot h_{\mu\nu}
+\dot A\bar g_{\mu\nu}\dot h)=-4\pi G_5(T_{\mu\nu}-\ft13\bar g_{\mu\nu}T)
\delta(y),\cr
&\bar g^{\nu\lambda}
(\nabla_\rho\dot h_{\nu\lambda}-\nabla_\lambda\dot h_{\nu\rho})=0,\cr
&\ddot h+2\dot A{\dot h}=\ft43\pi G_5 T\delta(y).\cr}
\numbereq\name{\eqriccota}
$$
The same set of equations was also given in
\ref{\kkr}{A. Karch, E. Katz and L. Randall,
\rtitle{Absence of a VVDZ discontinuity in $AdS_{AdS}$}
\jhep0112 016 (2001) [hep-th/0106261].}.
Here we have included the matter stress energy tensor, $T_{\mu\nu}\delta(y)$,
as a source.  This is in addition to the Karch-Randall brane tension itself,
which is accounted for by the kink in the warp factor.  By integrating the
equations (\eqriccota) across the brane, one would obtain junction conditions
relating the discontinuity of the normal derivative of the metric $h_{\mu\nu}$
to the matter sources on the brane.  Such conditions are of course
equivalent to the Israel matching conditions arising from the matter
on the brane.  These linearized gravity equations, (\eqriccota), will be
the starting point of our investigation.

Let us keep in mind the fact that the Gauss-normal form of the metric,
(\eqds), admits residual gauge transformations 
$$
\eqalign{
x^\mu&\to x^\mu+\phi^\mu(x)-{1\over k^2}\dot A\nabla^\mu \chi(x),\cr
y&\to y+\chi(x),}
\numbereq\name{\eqdiff}
$$
generated by the $x$-dependent functions $\phi^\mu(x)$ and $\chi(x)$.  The
transformation of the metric is given by
$$
h_{\mu\nu}\to h_{\mu\nu}+\nabla_\mu\phi_\nu+\nabla_\nu\phi_\mu
+{2\over k^2}\dot A(-\nabla_\mu\nabla_\nu\chi+k^2\bar g_{\mu\nu}\chi).
\numbereq\name{\eqgauge}
$$
While these transformations leave the Gauss-normal form of the metric
invariant, the shift $y\to y+\chi(x)$ in (\eqdiff) is a brane-bending
transformation
\lref{\gt}{J. Garriga and T. Tanaka,
\rtitle{Gravity in the brane-world}
Phys.\ Rev.\ Lett.\  {\bf 84}, 2778 (2000) [hep-th/9911055].}%
\lref{\gkr}{S.B. Giddings, E. Katz and L. Randall,
\rtitle{Linearized gravity in brane backgrounds}
JHEP {\bf 0003}, 023 (2000) [hep-th/0002091].}%
[\gt,\gkr],
since it moves the location of the brane, $y=0$, to $y=\chi$.  The gauge
transformation (and brane bending) function $\chi(x)$ is closely related to
the radion.

We now consider the requirement of conservation of the brane stress tensor.
In particular, we are interested in the bulk divergence $T_{MN}^{(s)\,;N}$.
This may be computed to first order in $h_{\mu\nu}$.  The result is
$$\eqalign{T_{\mu N}^{(s)\,;N}&=0,\cr
T_{4N}^{(s)\,;N}&\propto 2(4+h)\dot A(y)\delta(y-y_s)+\dot h\delta(y-y_s).}
\numbereq\name{\eqdiv4}
$$
Here, $y_s$ is the location of the brane, {\it i.e.}~$y_s=0$ for $s=0$ (the
physical brane) and $y_s=y_1$ for $s=1$ (the regulator brane).  Since 
$\dot A(y)\propto \sgn(y-y_s)$ and $\sgn(z)\delta(z)$ may
always be regulated to zero, the first term in (\eqdiv4) is unimportant,
and we are left with the second term.  The vanishing of the divergence
then demands $\dot h(y_s)=0$.  Thus conservation of energy-momentum
demands Newmann boundary conditions in general.  However, Dirichlet boundary
conditions are also permissible for a traceless mode, $h=0$.

\newsection The transverse-traceless modes.%
Now that we have set up the linearized gravity equations, we will reexamine
the braneworld eigenmode equations paying particular attention to boundary
conditions.  Since the radion mode is of particular interest, and since
its treatment demands care, it will be deferred to the following section.
Here we focus on the quasi-zero mode graviton and the corresponding
Kaluza-Klein tower.

Based on energy momentum conservation, we always impose Newmann boundary
conditions on the physical brane.  On the other hand, we consider both
Newmann and Dirichlet boundary conditions on the regulator brane.  We
begin with the two brane scenario and then show how the one brane case
may be recovered (for either boundary condition on the regulator brane)
by pushing the second brane to infinity.

Since the $y$-dependence of the trace and of the longitudinal
components of the metric can be readily determined by
the second and third equations of (\eqriccota), we shall
focus only on the transverse-traceless components. By writing
$h_{\mu\nu}(x, y)={\chi_{\mu\nu}}(x)\psi(y)$ where $\chi=0$ and
$\nabla^\mu\chi_{\mu\nu}=0$, we find that $\chi_{\mu\nu}$ satisfies the
transverse-traceless spin-2 equation
$$
{\nabla^2}{\chi_{\mu\nu}}+(2k^2-\mu^2)\chi_{\mu\nu}=0,
\numbereq\name{\eqxios}
$$
while $\psi$ obeys the eigenvalue equation
$$
{\ddot\psi}+4{\dot A}{\dot\psi}={\mu^2}e^{-2A}\psi.
\numbereq\name{\eqweir}
$$
The eigenvalue $\mu^2$ is identified as the four-dimensional mass of the
mode.  From (\eqfluc), the off-shell action of this mode reads
$$
S=\fft{K}2||\psi||^2{\int}d^{4}x{\sqrt{-\bar g}}h^{\mu\nu}
({\nabla^2}-\mu^2+2k^2)h_{\mu\nu}
+{1\over 2}\int_{y=0} d^4x\sqrt{-\bar g}h^{\mu\nu}T_{\mu\nu},
\numbereq\name{\eqpios}
$$
where the norm of $\psi(y)$ given by
$$
||\psi||^2={\int_{0}^{y_1}}dy\,e^{2A}\psi^2.
\numbereq\name{\eqveritas}
$$
Note that this is also the norm appropriate to the Sturm-Liouville problem
defined by (\eqweir). In what follows, we shall set the value of
$\psi$ to be finite on the physical brane, so that a diverging norm in
the limit $y_1 \to \infty$ would imply zero susceptibility to
matter on the physical brane (so that the corresponding mode decouples).

We now have a choice of boundary conditions for the regulator brane.  Let
us first consider imposing Newmann boundary conditions, $\dot\psi(y_1)=0$.
In this case, there exists a trivial zero mode, $\mu^2=0$, with eigenfunction
$\psi=1$ and corresponding norm
$$
||\psi||^2={k^2\over {4{\kappa}^3}}[{\kappa}y_1-
{1\over 2}{\sinh{2{\kappa}(y_1-\bar y)}}-{1\over 2}
{\sinh{{\kappa}\bar y}}].
\numbereq\name{\eqbyron}
$$
For $\mu^2 \ne 0$, we introduce the variable $\zeta=\tanh \kappa(y-\bar y)$
and the function $\phi=\psi/(1-\zeta^2)$. The mode equation, (\eqweir), is
then converted into an associated Legendre equation for $\phi$
$$
(1-\zeta^2){d^2\phi\over d{\zeta}^2}
-2\zeta{d\phi\over d{\zeta}}+\left(l(l+1)-{m^2\over {1-\zeta^2}}\right)\phi=0,
\numbereq\name{\eqeric}
$$
with $l=E_0-2$ and $m=2$.  Here, we have introduced $E_0$, which is the
lowest energy eigenvalue for the massive spin-2 representation $D(E_0,s=2)$
of the AdS$_4$ isometry group SO(2,3).  The value of $E_0$ is related to
the four-dimensional mass according to
$$
\mu^2=E_0(E_0-3)k^2=(l+2)(l-1)k^2.
\numbereq\name{\eqmathra}
$$
The stability condition of the AdS$_4$ brane, $\mu^2>-{9\over 4}k^2$, rules 
out complex values for $l$ and the symmetry of (\eqeric) with respect to 
$l\to -(l+1)$ narrows our scope down to $l>-{1\over 2}$. The case of $l=1$, 
which gives rise to $\mu^2=0$ has just been discussed and the case with 
$l=0$ will be postponed to the next section. 

For a non-compact $y$ interval, the range of the new variable $\zeta$ would
lie between the regular singular points $-1$ and $1$. Furthermore, the
normalizability condition dictates $l=2,3,4,\ldots$. However, the presence
of both the physical and regulator brane cuts off this range, so that
$\zeta$ is constrained to lie in the interval
$-1+2\xi <\zeta< 1-2\eta$, where
$$
{\xi}={1\over 2}(1-{\tanh{{\kappa}\bar y}}), \qquad
\eta={1\over 2}(1-{\tanh{{\kappa}(y_1-\bar y)}}).
\numbereq\name{\eqtilion}
$$
For $\xi \ll 1$ and $\eta \ll 1$ the corresponding value of $l$ will be 
shifted slightly away from the above integer sequence and a quasi zero mode is 
expected to emerge with $l\simeq 1$. 
For non-integer $l$, the most general solution to 
the associated Legendre equation, (\eqeric),
is a linear combination of $P_l^m(\zeta)$ and $P_l^m(-\zeta)$
with $m=2$, {\it i.e.}
$$
\psi(\zeta)=(1-\zeta^2)[c_1P_l^2(\zeta)+c_2P_l^2(-\zeta)].
\numbereq\name{\eqbravo}
$$

The quantization condition on $l$ (and hence the Kaluza-Klein mass spectrum)
arises from imposing Newmann boundary conditions, which are simply
${d\psi/d\zeta}\vert_{\zeta=-1+2\xi}=0$ for the physical brane and
${d\psi/d\zeta}\vert_{\zeta=1-2\eta}=0$ for the regulator brane.  Using the
representation of the associated Legendre functions in terms of
hypergeometric functions
$$
P_{l}^{\,2}(\zeta)={\Gamma(l+3)\over 8\,\Gamma(l-1)}(1-\zeta^2)
\,{}_2F_1(2-l,l+3;3;\ft12(1-\zeta)),
\numbereq\name{\eqbrink}
$$
we find that the wavefunction and its derivative on the
physical brane are given by
$$
\eqalign{
\psi(-1+2\xi)&\simeq (-)^{l_0}\lbrace -4{\delta}l+2(l-1)(l+2) [-2\delta l\xi+
l(l+1)(1+(-)^{l_0}c_2)\xi^2] \rbrace,\cr
\fft{d}{d\zeta}\psi(-1+2\xi)&\simeq 2(-)^{l_0}(l-1)(l+2)[-{\delta}l
+l(l+1)(1+(-)^{l_0}c_2)\xi],}
\numbereq\name{\eqquaternion} 
$$
where we have set $c_1=1$ and $l=l_0+{\delta}l$, with $l_0$
an integer and $\delta l<<1$. The analogous expressions on the regulator
brane may be obtained from (\eqquaternion) by symmetry
$$
\eqalign{
\psi(1-2\eta) &\simeq \lbrace -4(-)^{l_0}c_2{\delta}l+2(l-1)(l+2)[
-2(-)^{l_0}c_2{\delta}l\eta+l(l+1)(1+(-)^{l_0}c_2)\eta^2] \rbrace,\cr
\fft{d}{d\zeta}\psi(1-2\eta)&\simeq -2(l-1)(l+2)[(-)^{l_0+1}c_2{\delta}l
+l(l+1)(1+(-)^{l_0}c_2)\eta].}
\numbereq\name{\eqcomplex} 
$$
As a result, the Newmann--Newmann (NN) boundary conditions yield the relations
$$
\eqalign{
-{\delta}l+l(l+1)(1+(-)^{l_0}c_2)\xi&=0\cr
-(-)^{l_0}c_2{\delta}l+l(l+1)(1+(-)^{l_0}c_2)\eta&=0\cr}
\numbereq\name{\eqnew}
$$
which are satisfied for
$$
{\delta}l=l_0(l_0+1)(\xi+\eta), \qquad c_2=(-)^{l_0}{{\eta}\over
{\xi}}.
\numbereq\name{\eqold}
$$
Thus the Kaluza-Klein graviton spectrum in the KR1 model with NN
boundary conditions is given by
$$
E_0=\cases{3,&zero mode graviton,\cr
l_0+2+l_0(l_0+1)(\xi+\eta)+\cdots,&$l_0=1,2,\ldots$}
\numbereq\name{\specnn}
$$
where $\xi$ and $\eta$, as given in (\eqtilion), parametrize the locations
of the branes.  In this scenario, both the true zero mode graviton
($\psi(y)=1$, $l=1$) and the $l_0=1$ quasi-zero mode are present in the 
spectrum.

If on the other hand we demand that the solution satisfies Dirichlet
conditions on the regulator brane (ND boundary conditions) we would obtain
$$
-{\delta}l+l(l+1)(1+(-)^{l_0}c_2)\xi=0, \qquad c_2=0
\numbereq\name{\eqcold}
$$
which is solved by
$$
{\delta}l=l_0(l_0+1)\xi, \qquad c_2=0.
\numbereq\name{\eqorion}
$$
In this ND case, the trivial zero mode, $\psi(y)=1$, is no longer permissible.
As a result, the spectrum is simply
$$
E_0=l_0+2+l_0(l_0+1)\xi+\cdots,\qquad l_0=1,2,\ldots
\numbereq\name{\specnd}
$$

While the true zero mode graviton is only present in the NN case, the
massive Kaluza-Klein spectra, (\specnn) and (\specnd) are similar.  Taking
the KR2 (single brane) limit, $y_1 \to \infty$, we see that the trivial
NN zero mode graviton becomes non-normalizable and decouples. Furthermore,
the Kaluza-Klein spectra, (\specnn) and (\specnd), converge and yield a
single smooth limit which reproduces the KR2 results of
\lref{\glr}{I. Giannakis, J.T. Liu and H.C. Ren,
\rtitle{Linearized gravity in the Karch-Randall braneworld}
Nucl.\ Phys.\ B {\bf 654}, 197 (2003) [hep-th/0211196].}%
[\karch,\glr].
In particular, both Kaluza-Klein towers limit to that of the ND case,
(\specnd), and the resulting quasi-zero mode graviton ($l_0=1$) mass
$$
\mu^2 \simeq {3\over 2}{k^4\over {\kappa^2}},
\numbereq\name{\eqwot}
$$
is recovered in the limit, starting from the KR1 model with either NN or
ND boundary conditions.

\newsection The radion mode.%
We have seen in the last section that the requirement of imposing
homogeneous boundary conditions on both branes (whether NN or ND) gives
rise to an eigenvalue problem and hence a discrete Kaluza-Klein mass
spectrum.  For other values of the mass, $\mu^2$, it is impossible to
satisfy both homogeneous boundary conditions simultaneously.  This argument
that compact extra dimensions give rise to a discrete mass spectrum with
a gap is of course just the standard one.

There is a slight subtlety to keep in mind, however, and that has to deal
with the fact that the boundary conditions are imposed at separate locations.
In particular, since a single Gauss-normal coordinate patch cannot cover
both branes simultaneously, we must impose the boundary conditions on
separate coordinate patches.  Since the physical bulk is unique and unaffected
by choice of coordinates, this means in practice that the two simultaneous
boundary conditions may be relaxed up to a gauge transformation with respect
to each other.  The result will not contribute to the physics of the two
branes.

This issue of brane bending does not have to be directly addressed in the
case of the transverse-traceless modes in vacua, as considered above.
However, for other modes, it is often unavoidable.  In fact, this is
precisely the case for the
Karch-Randall radion, which has $\mu^2=-2k^2$, or equivalently $l=0$.
For $\mu^2=-2k^2$, it is easy to see that the mode equation (\eqweir)
admits two linearly independent solutions,
$$
\psi_1=\dot A, \qquad \psi_2=(\kappa-\dot A)^2.
\numbereq\name{\eqwave1}
$$ 
Furthermore, it is important to note that $\psi_1$ matches the $y$
profile of the brane bending gauge transformation (\eqdiff) parametrized
by $\chi(x)$.  In this sense, $\psi_1$ is at least locally unphysical,
although care may have to be taken in extending the transformation into
multiple coordinate patches.  Note, however, that brane bending in isolation
is unphysical, as it only corresponds to choosing different coordinates
with which to describe the brane.  On the other hand, the second solution,
$\psi_2$, cannot be gauged away, and is always physical provided the solution
can be made to satisfy appropriate boundary conditions.

For the KR1 radion mode, the solutions $\psi_1$ and $\psi_2$ must be
combined to satisfy the appropriate boundary conditions.  As we have seen,
we may consider either NN or ND boundary conditions.  However, in order to
describe the solution in the neighborhood of both branes, we will need to
introduce two coordinate patches.  The first one will be straight
with respect to the physical brane (so that the location of the physical
brane is at $y=0$) while the other one will be straight with respect to
the regulator brane.  The combined solution for the radion in these
two coordinate systems will be denoted by the pair $\psi_{\rm phy}$ and
$\psi_{\rm reg}$, where
$$
\eqalign{\psi_{\rm phy}(\zeta) = c_1\zeta + c_2(1-\zeta^2),\cr
\psi_{\rm reg}(\zeta) = d_1\zeta + d_2(1-\zeta^2).}
\numbereq
$$
Although these functions are defined in separate coordinate patches, we
nevertheless use the same coordinate notation, $\zeta$, for convenience.
One should, however, keep in mind that these are not directly comparable
without first performing the appropriate gauge transformation.

While $\psi_{\rm phy}$ and $\psi_{\rm reg}$ provide the appropriate $y$
profiles of the radion, the
wavefunction on the brane may be denoted $\chi(x)$.  In fact, a residual
coordinate transformation of the form (\eqdiff) must necessarily relate
the two solutions $\psi_{\rm phy}$ and $\psi_{\rm reg}$. As a result the
metric fluctuations corresponding to the radion mode can be written as 
$$
h_{\mu\nu}=(-{\nabla_\mu}{\nabla_\nu}+k^2{\bar g}_{\mu\nu})\chi\psi.
\numbereq\name{\eqradion}
$$
In addition, the transverse-traceless condition yields the simple
equation of motion
$$
(-{\nabla^2}+4k^2)\chi=0,
\numbereq\name{\eqvart}
$$
appropriate to an $E_0=4$ scalar.
 
Let us now impose appropriate boundary conditions on $\psi_{\rm phy}$
and $\psi_{\rm reg}$.  Focusing on the NN case, for branes located at
$\zeta=-1+2\xi$ and $\zeta=1-2\eta$ (in their respective coordinate
patches) we demand
$$
\fft{d}{d\zeta}\psi_{\rm phy}(-1+2\xi)=0,\qquad
\fft{d}{d\zeta}\psi_{\rm reg}(1-2\eta)=0.
\numbereq\name{\bceqns}
$$
This is easily solved to obtain $c_1=4(1-\xi)c_2$ and $d_1=4d_2\eta$.
Furthermore, since the $\psi_2$ component of the wavefunction is unaffected
by the brane bending transformation, we may simultaneously choose $c_2=d_2=1$.
Then the radion profile takes the form
$$
\psi_{\rm phy}(\zeta)=4(1-\xi)\zeta+(1-\zeta)^2,
\numbereq\name{\eqwrist}
$$
on the physical brane, and
$$
\psi_{\rm reg}^{\rm N}(\zeta)=4\eta\zeta+(1-\zeta)^2,
\numbereq\name{\eqcarpet}
$$
on the regulator brane.  Since $\psi_{\rm phy}\ne\psi_{\rm reg}$ (except
for the unreasonable case $\xi+\eta=1$ since both parameters are assumed
small), this demonstrates that it is impossible to satisfy both boundary
conditions simultaneously on a single coordinate patch.

So far, we have been working in Gauss-normal coordinates.  However,
to examine the off-shell radion action, it is advantageous to transform
the metric with a radion fluctuation, (\eqradion), to the generic 
form 
$$
ds^2=(e^{2A}+f)\bar g_{\mu\nu}dx^\mu dx^\nu+(1-2e^{-2A}f)dy^2,
\numbereq\name{\eqgene}
$$
which is now straight with respect to both branes. In order to reach this
form of the metric from the `physical' Gauss-normal patch, we apply the
coordinate transformation 
$$
\eqalign{
x^\mu&\to x^\mu+\epsilon_{\rm phy}^\mu(x,y)\cr
y&\to y+u_{\rm phy}(x,y)}
\numbereq\name{\eqphy}
$$
to (\eqwrist), where
$$
u_{\rm phy}(x,y)={k^2\over\kappa^2}[\dot A(0)-\dot A(y)]\chi(x),\qquad
\dot\epsilon_{\rm phy}^\mu(x,y)=-e^{-2A}\nabla^\mu u_{\rm phy}.
\numbereq\name{\eqtransphy}
$$
Similarly, transforming (\eqcarpet) from the `regulator' Gauss-normal
patch may be accomplished by taking
$$
\eqalign{
x^\mu&\to x^\mu+\epsilon_{\rm reg}^\mu(x,y)\cr
y&\to y+u_{\rm reg}(x,y)}
\numbereq\name{\eqreg}
$$ 
where
$$
u_{\rm reg}(x,y)={k^2\over\kappa^2}[\dot A(y_1)-\dot A(y)]\chi(x),\qquad
\dot\epsilon_{\rm reg}^\mu(x,y)=-e^{-2A}\nabla^\mu u_{\rm reg}.
\numbereq\name{\eqtransreg}
$$
Note that the transformation functions vanish on the appropriate branes,
$u_{\rm phy}(x,0)=u_{\rm reg}(x,y_1)=0$, and furthermore the behavior of
the metric (\eqgene) is governed by $f={k^4\over \kappa^2}\chi$ for both
cases.  The radion action is then given by
$$
S_{\rm radion}={3\over {32\pi{G_5}}}||\psi_{\rm radion}||^2
\int d^4x\sqrt{-\bar g}f(\nabla^2-4k^2)f
+{1\over 2}\int_{y=0}d^4x\sqrt{-\bar g}f\bar g^{\mu\nu}T_{\mu\nu},
\numbereq\name{\eqact}
$$
in agreement with that in [\chacko]
and
\ref{\papazoglou}{A. Papazoglou,
\rtitle{Dilaton Tadpoles and Mass in Warped Models}
Phys. Lett. {\bf B505}, 231 (2001) [hep-th/0102015].}.
Here, the radion norm is 
$$
||\psi_{\rm radion}||^2=\int_0^{y_1}dy\,e^{-2A},
\numbereq\name{\eqradn}
$$
and remains finite in the KR2 limit, $y_1\to\infty$.  As a result, we see
that the Karch-Randall radion survives with finite action, even in the
limit when the second brane is pushed off to the boundary of AdS$_5$.

We now turn to the case of ND boundary conditions.  For Dirichlet boundary
conditions on the second brane, the second condition of (\bceqns) is
replaced by $\psi_{\rm reg}(1-2\eta)=0$.  As a result, the corresponding
radion wavefunction takes the form
$$
\psi_{\rm reg}^{\rm D}(\zeta)=-{4{\eta}^2\over {1-2\eta}}\zeta
+(1-\zeta)^2.
\numbereq\name{\eqvirial}
$$
Just as in the NN case, this is inequivalent to $\psi_{\rm phy}$ of
(\eqwrist).  So again a brane bending transformation is unavoidable.
However, we observe that in the KR2 limit, $\eta \to 0$, the wavefunction
$\psi_{\rm reg}\to (1-\zeta)^2$ is the same regardless of NN or ND boundary
conditions.

While the coordinate transformation that brings the radion fluctuation
$\psi_{\rm phy}$ on the physical patch into the form (\eqgene) remains the
same as (\eqphy), the one on the regulator patch for $\psi_{\rm reg}^{\rm D}$
instead takes the form
$$
u_{\rm reg}^{\rm D}(x,y)=-{k^2\over\kappa^2}\dot A(y)\chi(x)
+{k^4\over 2\kappa^2 \dot A(y_1)}\Big[{2\kappa^2\over k^2}
-e^{-2A(y_1)}\Big]\chi(x).
\numbereq\name{\eqregp}
$$ 
Note that this transformation does not vanish on the regulator brane
$$u_{\rm reg}^{\rm D}(x,y_1)={k^4\over 2\kappa^2}
{e^{-2A(y_1)}\over\dot A(y_1)}\chi(x).
\numbereq\name{\eqtransregp}
$$
As a result, even the generic form of the metric, (\eqgene), is incapable
of describing two straight branes, except in the limit $y_1\to\infty$.

Note that the radion corresponds to $l=0$ exactly.  However, one could
ask whether any modes exist for $l$ close to zero which satisfies both
boundary conditions simultaneously (thus obviating the need of a brane
bending gauge transformation).  We demonstrate in the appendix that such
hypothetical modes do not exist.

\newsection Discussion.%
For a two brane scenario, whether RS1 or KR1, the radion mode is clearly
normalizable (as any mode would be on a compact space), and has a physical
interpretation as a modulus for the distance between the branes.  For this
reason, it is not surprising that it is a scalar mode that is closely related
to the brane bending gauge transformation and which couples universally to
the trace of the stress tensor on the brane.

Since the radion is connected to the separation of two branes, in the
limit when the regulator brane is pushed off to
infinity, the fate of the radion is closely tied to that of the brane.  As
shown in the previous section, the KR1 radion survives this limit (for
either NN or ND boundary conditions).  At the same time, the second brane
has not disappeared from view, but remains attached to the boundary
of AdS$_5$.  For the RS1 model, on the other hand, it may be argued that
since the second brane is pushed into a Cauchy horizon (so that its
presence can no longer be felt), the corresponding radion must disappear from
the spectrum in this limit.

This may be seen explicitly in the above analysis by considering the
Randall-Sundrum limit of the Karch-Randall model, $k\to0$.  To bring the
radion wavefunction $h_{\mu\nu}=(-{\nabla_\mu}{\nabla_\nu}+
k^2{\ov g}_{\mu\nu}){\chi}{\psi}$, with $\psi=\psi_{\rm phy}$,
to the familiar form [\charmousis] in this limit, let us first
perform a gauge transformation
$$
h_{\mu\nu}^{\prime}=h_{\mu\nu}+{k^2\over {\kappa^2}}
(1+{k^2\over 4{\kappa^2}}){\nabla_\mu}
{\nabla_\nu}\chi.
\numbereq\name{\eqpergolesi}
$$
Then we find that
$$
\lim_{k \to 0}{4{\kappa^4}\over k^4}h_{\mu\nu}^{\prime}=
(2e^{2{\kappa}y}-e^{4{\kappa}y}){\partial_\mu}{\partial_\nu}
\chi+4{\kappa^2}\eta_{\mu\nu}\chi.
\numbereq\name{\eqviros}
$$
Similarly we find for $\psi=\psi_{\rm reg}^{N}$ that
$$
\eqalign{
\lim_{k \to 0}{4{\kappa^4}\over k^4}\left(h_{\mu\nu}
+{k^2\over {\kappa^2}}e^{2{\kappa}y_1}\left[1+{k^2\over 4{\kappa^2}}
(2-e^{2{\kappa}y_1})\right] {\nabla_\mu}{\nabla_\nu}{\chi}\right)&\cr
&\kern-4cm
=(2e^{2{\kappa}y+2{\kappa}y_1}-e^{4{\kappa}y}){\partial_\mu}{\partial_\nu}
\chi+4{\kappa^2}e^{2{\kappa}y_1}\eta_{\mu\nu}\chi. \cr}
\numbereq\name{\eqvirko}
$$
where $\chi$ satisfies the massless flat-space scalar equation
$\nabla^2\chi=0$.  The right hand side of (\eqviros) is in fact the RS1
radion mode.  Upon transforming this to the generic coordinate system
with two straight branes, (\eqgene), and noting that $e^{2A}=e^{-2\kappa y}$,
we identify the same form of the off-shell radion action as (\eqact), however
with a norm that diverges like $e^{2{\kappa}y_1}$ as $y_1\to\infty$.

In particular, in the Randall-Sundrum limit, the normalization integral
pertaining to the Sturm-Liouville problem, (\eqveritas), 
becomes (up to an overall constant)
$$
||\psi||^2={\int_{0}^{y_1}}dy\,e^{-2{\kappa}y}\psi^2,
\numbereq\name{\ebasano}
$$
and hence the RS1 radion (\eqpergolesi) is not normalizable in
the RS2 limit, $y_1 \to \infty$. Note that the dominant non-normalizable
term, $e^{4{\kappa}y}$, cannot be gauged away through a brane bending
transformation. Therefore the radion will decouple in the RS2 limit,
unlike for the Karch-Randall model with a single brane. 

One may yet worry that the limit is not smooth, so that the KR2 model is
physically distinct from the KR1 model where the second brane is pushed to
infinity.  However, in order to make this claim, one would need a well
defined KR2 model to start with, in order to compare with the limiting
procedure of KR1.  In this case, starting with say the KR2 model, one
would have to impose boundary conditions at infinity so that the
modes remain normalizable.  Although such boundary conditions are not
straightforwardly classified as Dirichlet or Newmann, we have seen that
the appropriate KR1 modes have well defined and unique asymptotic
limits, regardless of NN or ND conditions.  As a result, the limiting
procedure, $y_1\to\infty$, applied to the KR1 model will yield normalizable
wavefunctions identical to those obtained by solving the KR2 system
directly.  Hence this limit is well defined, at least for all Karch-Randall
modes except the massless KR1 graviton, which becomes non-normalizable
in the limit.

The absence of a van Dam-Veltman-Zakharov discontinuity [\vdv,\zak]
for massive gravity in an AdS background [\Kogan,\Porrati] points to a
smooth critical tension ({\it i.e.} RS2) limit for the KR2 braneworld.
However, the results of [\Kogan,\Porrati] do not take the radion into
account.  Thus it is also necessary to see that the KR2 radion smoothly
decouples from the matter source on the brane to ensure a smooth critical
tension limit.  Of course, the recovery of the full RS2 solution from the
KR2 one for a point source on the brane [\glr], the smoothness of the
KR1 $\to$ KR2 limit, and the KR1 $\to$ RS1 $\to$ RS2 limit all indicate
that this ought to be the case.

To prove that the radion indeed smoothly decouples, we first note that
the KR2 radion action is given by (\eqact) in the $y_1\to\infty$ limit.
In this case, the coupling strength of the KR2 radion to matter is 
inversely proportional to the norm $||\psi_{\rm radion}||$, given by
$$||\psi_{\rm radion}||^2=\int_0^\infty dye^{-2A}={\kappa\over k^2}
(1+{\rm tanh}\kappa \bar y)={\kappa\over k^2}
\Big(1+\sqrt{1-{k^2\over\kappa^2}}\Big).
\numbereq\name{\eqnormKR2}
$$
In the RS2 limit, $k\to 0$, the norm diverges, and hence the radion coupling
vanishes.  Since both the massive graviton and the radion give rise to
smooth limits, we conclude that the van Dam-Veltman-Zakharov discontinuity
is absent from the KR2 model, in accord with expectations.

A physical understanding of the origin of the KR2 radion may be seen without
resorting to the limiting case of the KR1 model.  We first recall that for
the RS2 scenario, the wrap factor, given by (\eqRS), is monotonically
and uniformly decreasing as one moves into the bulk.  So in a sense, there
is no preferred location off the brane.  On the other hand, for the KR2
model, the warp factor of (\eqKR) first decreases, but then turns around and
increases without bound as one moves into the bulk.  The warp factor reaches
a minimum at $y=\bar y$, which may be viewed as a preferred location in the
bulk (based upon the embedding of AdS$_4$ in AdS$_5$).  This is especially
true in applying the holographic interpretation to the KR2 model, where
$\bar y$ is the locus separating multiple holographic domains
\lref{\porhol}{M. Porrati,
\rtitle{Mass and gauge invariance. IV: Holography for the Karch-Randall
model}
Phys. Rev. {\bf 65}, 044015 (2002) [hep-th/0109017].}%
\lref{\bora}{R. Bousso and L. Randall,
\rtitle{Holographic domains of anti-de Sitter space}
JHEP {\bf 0204}, 057 (2002) [hep-th/0112080].}%
[\porhol,\bora].
The radion is then connected to the transverse distance between the brane
(located at $y=0$) and the boundary of its holographic domain (located at
$y=\bar y$).  In fact, a four-dimensional CFT analysis of the holographic
dual of the KR2 model has previously shown hints of a radion in the
spectrum [\Duffkr].  It would be worthwhile to revisit this analysis and
to extract the appropriate radion behavior from the two-point function of
the stress tensor of the CFT dual.

\noindent{\twelvebf Acknowledgments.}

This research grew out of a series of discussions with A. Karch. The 
authors are very grateful for his enlightening comments and valuable 
suggestions. They are also indebted to Z. Chacko for helpful communications.  
This work was supported in part by the US Department of Energy under
grants DE-FG02-91ER40651-TASKB and DE-FG02-95ER40899.
                                                                                
\newappendix Appendix.%
In Section 4, we hare argued that the KR braneworld does not admit any
additional modes for $l$ near zero, and hence that we have identified the
complete spectrum of the system.  Here, we demonstrate explicitly the
absence of such a mode with $l \simeq 0$ that satisfies Neumann boundary
conditions on both branes.

Starting with (\eqeric), we see that the mode equation for $l \simeq 0$
can be written as
$$
{{d^2\psi}\over {d{\zeta^2}}}+{2{\zeta}\over {1-{\zeta^2}}}
{{d\psi}\over d{\zeta}}
-{2\over {1-{\zeta^2}}}{\psi}=-{{l(l+1)}\over {1-{\zeta^2}}}{\psi}
\simeq -{l\over {1-{\zeta^2}}}{\psi^{(0)}},
\numbereq\name{\eqtakis}
$$
where $\psi^{(0)}$ is the solution to this same equation, however
with $l=0$.  The first order solution can be written $\psi^{(1)}=
\psi^{(0)}+d_1{\zeta}+d_2(1-\zeta)^2$ with the coefficients
$d_1$ and $d_2$ given by
$$
\eqalign{
d_1&=l{\int^\zeta}d{\xi}{{\psi^{(0)}(1-\xi)^2}\over {(1-\xi^2)W}}
=-l{\int^\zeta}d{\xi}{{\psi^{(0)}}\over {(1+\xi^2)}},\cr
d_2&=-l{\int^\zeta}d{\xi}{{\psi^{(0)}\xi}\over {(1-\xi^2)W}}
=l{\int^\zeta}d{\xi}{{\psi^{(0)}}\over {(1-\xi^2)^2}},}
\numbereq\name{\ewarp}
$$
and $W(\zeta, (1-\zeta)^2)=-1+\zeta^2$ is the Wronskian.  We now examine
the two linearly independent solutions of (\eqtakis).  Starting with
the zeroth order solution $\psi^{(0)}=\zeta$, we find
$$
d_1=-l\Big[{1\over {1+\zeta}}+{\ln(1+\zeta)}\Big], \qquad
d_2={l\over 2}\Big[{{\zeta}\over {1-{\zeta^2}}}-{1\over 2}
{\ln{{1+\zeta}\over{1-\zeta}}}\Big],
\numbereq\name{\eqwer}
$$
so that
$$
\psi^{(1)}=\zeta-{l\over 2}\Big[\zeta+{1\over 2}(1+\zeta)^2
{\ln(1+\zeta)}-{1\over 2}(1-\zeta)^2
{\ln(1-\zeta)}\Big].
\numbereq\name{\eqerion}
$$
Similarly, for $\psi^{(0)}=(1-\zeta)^2$, we find
$$
d_1=-l\Big[\zeta-{4\over {1+\zeta}}-4{\ln(1+\zeta)}\Big], \qquad
d_2={l}\Big[{1\over {1+\zeta}}+{\ln(1+\zeta)}\Big],
\numbereq\name{\eqweir}
$$
and
$$
\psi^{(1)}=(1-\zeta)^2-{l}\lbrack {\zeta^2}-1-\zeta-(1+\zeta)^2
{\ln(1+\zeta)}\rbrack.
\numbereq\name{\eqerifion}
$$

We now write the general solution as a linear combination of
even $\psi_{+}$ and odd $\psi_{-}$ combinations,
$\psi^{(1)}=c_{+}\psi_{+}+c_{-}\psi_{-}$, where
$$
\eqalign{
\psi_{+}&=1+\zeta^2+l\lbrack 1-\zeta^2+{1\over 2}(1+\zeta)^2
{\ln(1+\zeta)+{1\over 2}
(1-\zeta)^2{\ln(1-\zeta)}} \rbrack,\cr
\psi_{-}&=2\zeta+l\lbrack \zeta+{1\over 2}(1+\zeta)^2{\ln(1+\zeta)
-{1\over 2}(1-\zeta)^2{\ln(1-\zeta)}} \rbrack.}
\numbereq\name{\eqvital}
$$
Subsequently we demand that the radion wavefunction satisfies
Newmann boundary conditions both on the physical brane
located at $\zeta=-1+2\xi$ and on the regulator brane at
$\zeta=1-2\eta$. This requirement leads to the following
two equations for $c_{+}$ and $c_{-}$:
$$
\eqalign{
&[1-l(1+{\ln 2})]c_{-}+[-2+l(1-2{\ln 2})]c_{+}=0,\cr
&[1-l(1+{\ln 2})]c_{-}+[2-l(1-2{\ln 2})]c_{+}=0.}
\numbereq\name{\eqpion}
$$
Since the determinant of the coefficient matrix for this set of equations
is nonvanishing for $l\ll1$, no non-trivial solutions are possible.  This
demonstrates the absence of any additional modes near $l\simeq0$ compatible
with NN boundary conditions.

\immediate\closeout1
\bigbreak\bigskip

\line{\twelvebf References. \hfil}
\nobreak\medskip\vskip\parskip

\input refs

\vfil\end

\bye